\documentclass[twocolumn,british,aps,prd,superscriptaddress,preprintnumbers,floatfix,nofootinbib,
showkeys,showpacs]{revtex4-1}

\usepackage[latin9]{inputenc}
\usepackage{babel}
\usepackage{amsmath}
\usepackage{amssymb}
\usepackage{graphicx}
\usepackage{esint}

\makeatletter

\DeclareTextSymbolDefault{\textquotedbl}{T1}
\providecommand{\tabularnewline}{\\}

\usepackage{latexsym}
\usepackage{ulem}
\usepackage{epsfig}
\usepackage{epstopdf}
\usepackage{slashed}

\def\lsim{\mathrel{\raise.3ex\hbox{$<$\kern-.75em\lower1ex\hbox{$\sim$}}}}
\def\gsim{\mathrel{\raise.3ex\hbox{$>$\kern-.75em\lower1ex\hbox{$\sim$}}}}
\newcommand{\mpl}{m_\text{P}}

\newcommand{\expon}{e^{\frac{\varphi}{\sqrt{\alpha}\mpl}}}
\newcommand{\expontwoesp}{e^{2x}}

\newcommand{\esp}{\mathrm{ESP}}

\makeatother

\begin{document}
\global\long\def\expontwoesp{\mathrm{e}^{2x}}%
\global\long\def\esp{\mathrm{ESP}}%
\global\long\def\vac{\mathrm{vac}}%
\global\long\def\mpl{m_{\mathrm{Pl}}}%
\global\long\def\c{\mathrm{c}}%

\title{Quintessential inflation with a trap and axionic dark matter}
\author{Konstantinos Dimopoulos}
\email{konst.dimopoulos@lancaster.ac.uk}

\affiliation{Consortium for Fundamental Physics, Physics Department,Lancaster University,
Lancaster LA1 4YB, United Kingdom}
\author{Mindaugas Kar\v{c}iauskas}
\email{mindaugas.k@ucm.es}

\affiliation{Departamento de F\'isica Te\'orica and Instituto de F\'isica de
Part\'iculas y del Cosmos IPARCOS, Universidad Complutense de Madrid,
E-28040 Madrid, Spain}
\author{Charlotte Owen}
\email{c.owen@lancaster.ac.uk}

\affiliation{Consortium for Fundamental Physics, Physics Department,Lancaster University,
Lancaster LA1 4YB, United Kingdom}
\begin{abstract}
We study a new model of quintessential inflation which is inspired
by supergravity and string theory. The model features a kinetic pole,
which gives rise to the inflationary plateau, and a runaway quintessential
tail. We envisage a coupling between the inflaton and the Peccei-Quinn
(PQ) field which terminates the roll of the runaway inflaton and traps
the latter at an enhanced symmetry point (ESP), thereby breaking the
PQ symmetry. The kinetic density of the inflaton is transferred to
the newly created thermal bath of the hot big bang due to the decay
of PQ particles. The model successfully accounts for the observations
of inflation and dark energy {with natural values of the model parameters},
while also resolving the strong CP problem of QCD and generating axionic dark
matter, without isocurvature perturbations. Trapping the inflaton at the ESP
ensures that the model does not suffer from the infamous 5th force problem,
which typically plagues quintessence.
\end{abstract}
\maketitle

\section{Introduction}

By now we have most of the history of the Universe figured out. The
Hot Big Bang model covers pretty much the entire timeline, from a
few seconds after the original explosion until the present time, almost.
However, the edges of the story are still unclear, because observations
suggest that the Universe is undergoing accelerated expansion at very
early and very late times and such expansion cannot be part of the
hot big bang, where the Universe is filled with relativistic and
non-relativistic
matter only. General relativity dictates that accelerated expansion
can occur only if the Universe is filled by an exotic substance, with
negative enough pressure. For the early Universe, the required substance
is usually taken as one (or more) potentially dominated scalar field
and the accelerated expansion phase is cosmic inflation \cite{Starobinsky:1980te,Sato:1981ds,Kazanas:1980tx,Guth:1980zm}
and the scalar is called the inflaton field. Inflation sets the initial
conditions of the Hot Big Bang. For the late Universe however, the
simplest explanation is vacuum density, due to a non-zero value of
the cosmological constant~$\Lambda$.

The cosmological constant has to be there but what is its value? The
most natural choice is given by the Planck scale, because this is
the only scale in general relativity and also because this is the
cutoff scale of particle theory (beyond this the theory breaks down).
This however is at odds with nature. This cosmological constant problem
\cite{Weinberg:1988cp} predates the observations of recent accelerated
expansion. The way it used to be addressed is by assuming that the
cosmological constant is set to zero by some \textsl{unknown} symmetry.
Once late accelerated expansion was observed, many authors suggested
that this amounted to observing the true value of the cosmological
constant. The problem is that the associated vacuum density has to
be comparable to the density of the Universe today, which is about
120 orders of magnitude smaller than the ``natural'' value of $\Lambda$.
This has been called (by Lawrence Krauss) ``the worst fine-tuning
in physics''. To avoid this fine tuning, it was suggested that, as
in inflation, the current accelerated expansion is due to another
potentially dominated scalar field, called quintessence; the fifth
element after baryons, dark matter, photons and neutrinos
\cite{Ford:1987de,Wetterich:1987fm,Ratra:1987rm}.
It is important to point out that the quintessence proposal does not
solve the cosmological constant problem because $\Lambda$ is still
assumed to be zero due to some unknown symmetry.

Since they are both based on the same idea, it is natural to unify
the two in quintessential inflation \cite{Peebles:1998qn,Wetterich:2013wza}
(for a recent list of references see
\cite{Dimopoulos:2017zvq,Dimopoulos:2017tud}
and \cite{Hossain:2014zma,Geng:2015fla}), which considers that the
inflaton field survives until today and becomes quintessence. Apart
from being economic, quintessential inflation attempts to treat the
early and late accelerated expansion phases in a single theoretical
framework. Moreover, there are some practical advantages as well;
for example the initial conditions of quintessence are determined
by the inflationary attractor.

A scalar potential which satisfies all the requirements of inflation
and dark energy observations is hard to formulate because of the $\sim\,$110
orders of magnitude difference in energy density between the inflationary
energy scale and the energy scale of dark energy today. { To incorporate such
a large difference in energy density scales often requires a very curved scalar
potential \cite{Dimopoulos:2002hm}, which produces inflationary predictions
incompatible with the Planck satellite observations \cite{Ade:2015lrj}.
Moreover, after inflation, the field rolls down the incredibly steep potential
and gains so much kinetic energy that it is transposed many Planckian distance
in field space. This means the potential is at risk from UV corrections,
throwing the predictability of the theory into question and destabilising the
flatness of the potential.}

In this paper, to overcome the above problems, we start with an exponential
potential with a non-canonical kinetic term, featuring a pole at the
origin, which can be theoretically motivated e.g. in supergravity
theories. We utilise a field redefinition to regain canonical kinetic
terms and thereby transpose the pole to infinity. Doing so, introduces
a plateau into the potential, ensuring inflationary observables match
the observations of the Planck satellite (which favours a plateau
inflation model). The model naturally features a quintessential tail.
To ensure the validity of our setup, we stop the roll of the inflaton
field by trapping it at an enhanced symmetry point (ESP) before it
travels over a super-Planckian distance in field space. We demonstrate
in detail that, through this trapping, we can transform the kinetic
density of the field into the radiation density of the hot big bang,
reheating the Universe.

Reheating in quintessential inflation is a challenging issue, because
it cannot occur through inflaton decay (as is otherwise typical in
inflation) because the inflaton must survive until the present and
become quintessence. A number of reheating mechanisms have been put
forward, the most important of which are gravitational reheating
\cite{Ford:1986sy}\cite{Chun:2009yu},
instant preheating \cite{Felder:1998vq}\cite{Campos:2002yk}, curvaton
reheating \cite{Feng:2002nb}\cite{BuenoSanchez:2007jxm} and recently
non-minimal reheating \cite{Dimopoulos:2018wfg} (also called Ricci
reheating \cite{Opferkuch:2019zbd}). Reheating the Universe through
trapping the runaway inflaton is a novel mechanism, although the trapping
mechanism has been considered before in quintessential inflation \cite{BuenoSanchez:2006fhh}
\cite{BuenoSanchez:2006epu}, but only for being responsible for the
inflation part of the scenario as in trapped inflation \cite{Kofman:2004yc}.

In the spirit of economy, aligned with the philosophy behind quintessential
inflation, we consider that the ESP is due to a coupling of the inflaton
direction with the Peccei-Quinn field \cite{Peccei:1977ur,Peccei:1977hh},
so that after trapping, the Peccei-Quinn phase transition confines
the inflaton and generates a large inflaton mass such that there is
no threat of violation of the equivalence principle (5th force problem),
which typically is a problem with quintessence \cite{Carroll:1998zi,Wetterich:2004ff,Acharya:2018deu}.
The field remains trapped with non-zero potential density, which explains
the dark energy observations, while the theory also incorporates the
QCD axion, which can be the dark matter \cite{Baer:2014eja}.

We use natural units where %
\mbox{%
$c=\hbar=k_{{\rm B}}=1$%
} and %
\mbox{%
$8\pi G=m_{P}^{-2}$%
}, where %
\mbox{%
$m_{P}=2.43\times10^{18}\,$GeV%
} is the reduced Planck mass.

\section{The Model\label{sec:The-Model}}

We start with a Lagrangian density well motivated in both supergravity
and string theory\footnote{For example, 
 the K\"{a}hler potential for a string modulus $T$, is %
\mbox{%
$K/m_{{\rm Pl}}^{2}=-3\ln(T+\overline{T})=-3\ln(\sqrt 2\phi/m_{{\rm Pl}})$%
}, where %
\mbox{%
$T=\frac{1}{\sqrt{2}}(\phi+i\sigma)/m_{{\rm Pl}}$%
}, with %
\mbox{%
$\phi,\sigma\in\,$I$\!$R%
}. Then the kinetic term is given by
\[
\mathcal{L}_{\mathrm{kin}}=K_{T\bar{T}}\partial_{\mu}T\partial^{\mu}\bar{T}=\frac{3}{2}\left(\frac{\mpl}{\phi}\right)^2\left[\frac{1}{2}\left(\partial\phi\right)^{2}+\frac{1}{2}\left(\partial\sigma\right)^{2}\right]\,,
\]
where we considered %
\mbox{%
$T+\overline{T}=\sqrt{2}\phi/\mpl$%
} and the subscripts of the K\"{a}hler potential denote differentiation.
In our considerations we assume that the ESP lies at a minimum in
the direction of $\sigma$. {We also assume that $\sigma$ is heavy during
inflation so that there are no issues with excessive non-Gaussianity or 
isocurvature perturbations.} In this example, \mbox{$\alpha=3/2$}.}
with a perturbative and non-perturbative part: 
\begin{multline}
\mathcal{L}=\frac{\alpha}{2}\Big(\frac{\mpl}{\phi}\Big)^{2}(\partial\phi)^{2}+\frac{\left(\partial\chi\right)^{2}}{2}-V_{0}\,e^{-\kappa\phi/\mpl}\\
-V\left(\chi\right)-\frac{g^{2}}{2}\left(\phi-\phi_{\mathrm{ESP}}\right)^{2}\chi^{2}\,,\label{L-org}
\end{multline}
The $\chi$ field is taken to be the Peccei-Quinn field \cite{Peccei:1977ur,Peccei:1977hh}
associated with the U(1) Peccei-Quinn symmetry, whose Pseudo-Nambu-Godlstone
boson is the QCD axion, which is a prominent dark matter candidate
\cite{Baer:2014eja}. The order parameter $f$ is called the axion
decay constant. In this case, we have 
\begin{equation}
V\left(\chi\right)=\frac{\lambda}{4}\left(\chi^{2}-f^{2}\right)^{2}\,,\label{Vchi}
\end{equation}
\begin{equation}
\lambda\sim1\,,
\end{equation}
\begin{equation}
f\sim10^{12}\:\mathrm{GeV}\,.\label{f-val}
\end{equation}
In fact, the most likely range for the axion decay constant is %
\mbox{%
$10^{10}\:\mathrm{GeV}\lesssim f\lesssim10^{12}\:\mathrm{GeV}$%
}. The lower bound in this range comes from the SN1987A energy loss
rate, while the upper bound is required to avoid overproduction of
axions. However, this latter limit is dependent on assumptions regarding
the initial axion misalignment angle \cite{Baer:2014eja}. In this
paper we consider the estimate shown in Eq.~\eqref{f-val}, noting
that this choice does not make much difference in our results.

{It is important to point out that no bare cosmological constant (CC) is included
in the Lagrangian density in Eq.~\eqref{L-org}. This is because an unknown
symmetry is presumed to set the CC to zero, as was typically assumed even before
the observations of dark energy in order to overcome the infamous
``cosmological constant problem''. This problem is twofold: First, general
relativity may introduce a classical CC term in the Einstein-Hilbert action.
The only mass-scale in general relativity is due to Newton's gravitational
constant and is the Planck mass $8\pi G=1/m_{\rm Pl}^2$. But this cannot be the
mass-scale of the CC, so a new scale  must be included which is at most
$10^{-30}m_{\rm Pl}$. The problem is explaining why these two scales differ so
much. Second, quantum fields introduce a contribution to vacuum energy which
diverges and is presumed capped at the cutoff-scale of the theory, resulting in
a CC. At the moment this is at least the supersymmetry breaking scale
$>10^{-15}m_{\rm Pl}$. But this is unacceptable because observations suggest that
the CC is at most $10^{-30}m_{\rm Pl}$, for otherwise structure formation would be
inhibited. The ``solution'' to the CC problem (which predates the observation of
dark energy, as we mentioned) is to assume that some {\em unknown} mechanism
sets the CC to exactly zero. This is why there is no CC in our model. Then,
quintessence is used to explain the dark energy observations.

To assist our intuition,} we can make a field redefinition to regain a canonical
kinetic term with 
\begin{equation}
  \varphi=\sqrt{\alpha}\mpl\,\mathrm{ln}\left(\frac{\phi}{\mpl}\right)\,,
  \label{phivarphi}
\end{equation}
the Lagrangian density then becomes 
\begin{multline}
\mathcal{L}=\frac{(\partial\varphi)^{2}}{2}+\frac{(\partial\chi)^{2}}{2}-V_{0}\,\mathrm{exp}\Big(-\kappa\,e^{\frac{\varphi}{\sqrt{\alpha}\mpl}}\Big)-V\left(\chi\right)\\
-\frac{g^{2}\phi_{\esp}^{2}}{2}\left(e^{\frac{\varphi-\varphi_{\mathrm{ESP}}}{\sqrt{\alpha}\mpl}}-1\right)^{2}\chi^{2}\,.\label{eq:lagrangian_varphi}
\end{multline}

The inflaton potential now features a double exponential providing
a bridge between the vastly different energy scales of inflation and
dark energy. The model has a pole at $\phi=0$, which is transposed
to $\varphi=-\infty$ by the field redefinition, and a plateau appears
in the scalar field potential, providing the perfect location for
slow-roll inflation.\footnote{We consider only %
\mbox{%
$\phi>0$%
}.}

After inflation has completed, the field enters a period of kination
\cite{Spokoiny:1993kt,Joyce:1997fc}, where the dominant contribution
to its energy density is its kinetic energy, the field is oblivious
to the potential during this time and standard kination equations
can be used \citep{Dimopoulos:2017zvq}. After a brief period of kination,
the field crosses an enhanced symmetry point (ESP) at $\phi_{\esp}$
and, due to its coupling to the $\chi$ field, non-perturbative effects
will generate a sea of $\chi$ particles. The energy budget for the
particle production comes from the inflaton's kinetic energy and as
such the particle production promptly traps the inflaton at $\phi_{\esp}$.
If $\chi$ is coupled to the standard model, its subsequent decays reheat
the Universe. At this point the inflaton field's mass is primarily
dependent on $\chi$, which until this point has been held at $\chi=0$.
At a particular symmetry breaking scale (given by $f$) $\chi$ moves
to its non-zero vacuum expectation value (VEV), providing the inflaton
with a huge mass which acts to stop its motion. As such the $\varphi$
field is trapped at the ESP until late times when the residual energy
density can act as dark energy. We consider $\chi$ to be the radial
component of a complex field (the Peccei-Quinn field), whose angular
degree of freedom is an axion-like particle (ALP) (it can be the QCD
axion itself) which, whilst oscillating in the minimum of its potential,
can describe the dark matter in the Universe.

\section{Inflation}

The contribution to the inflaton potential from the coupling to the
$\chi$ field will not affect the inflationary dynamics because $\chi=0$
during inflation, due to the large mass it obtains from the interaction
term: 
\begin{equation}
m_{\chi}^{2}\left(\varphi\right)=g^{2}\phi_{\esp}^{2}\left(e^{\frac{\varphi-\varphi_{\mathrm{ESP}}}{\sqrt{\alpha}\mpl}}-1\right)^{2}\,,\label{eq:mass-squared-chi-1}
\end{equation}
where $\varphi\rightarrow-\infty$ during inflation. Hence, we have
the inflaton potential: 
\begin{equation}
V\left(\varphi\right)=V_{0}\,\mathrm{exp}\left(-\kappa\,e^{\frac{\varphi}{\sqrt{\alpha}\mpl}}\right)\,.\label{V-infl}
\end{equation}
From the expression above it is easy to compute the slow-roll parameters,
which are given by 
\begin{equation}
\epsilon\equiv\frac{\mpl^{2}}{2}\left(\frac{V_{\varphi}}{V}\right)^{2}=\frac{\kappa^{2}}{2\alpha}e^{\frac{2\varphi}{\sqrt{\alpha}\mpl}}\,,
\end{equation}
\begin{equation}
  \eta\equiv\frac{\mpl^{2}V_{\varphi\varphi}}{V}=\frac{\kappa}{\alpha}\expon\Big(\kappa\expon-1\Big)\,.
  \label{eta}
\end{equation}
Defining the end of inflation as $\epsilon=1$ leads us to 
\begin{equation}
\varphi_{\mathrm{end}}=\mpl\sqrt{\alpha}\,\mathrm{\ln}\,\frac{\sqrt{2\alpha}}{\kappa}\,,\label{phi-end}
\end{equation}
and 
\begin{equation}
\dot{\varphi}_{\mathrm{end}}\simeq\sqrt{\frac{2}{3}V_{\mathrm{end}}}\,,\label{dphi-end}
\end{equation}
where the latter was estimated by using the slow-roll equation $3H\dot{\varphi}\simeq-V_{\varphi}$
and 
\begin{equation}
V_{\mathrm{end}}\equiv V\left(\varphi_{\mathrm{end}}\right)=V_{0}\mathrm{e}^{-\sqrt{2\alpha}}\,.\label{Vend}
\end{equation}
We can use $\varphi_{\mathrm{end}}$ to find $\varphi$ when observable
scales first left the horizon%
\footnote{We can also use this result to compute the field excursion during
  inflation. For the non-canonical field we find
\begin{equation*}
  \frac{\phi_\mathrm{end}-\phi_{*}}{\mpl}=
\frac{\sqrt{2\alpha}}{\kappa}\frac{N_*}{N_*+\sqrt{\frac{\alpha}{2}}}
  \simeq  \frac{\sqrt{2\alpha}}{\kappa}\ll1\,,
\end{equation*}
which is well bellow the Planck range. Switching to the canonical field we get
$$
\varphi-\varphi_{\rm end}=-\sqrt\alpha m_P
\ln\left(1+\sqrt{\frac{2}{\alpha}}\,N_*\right)
\Rightarrow\Delta\varphi\simeq\sqrt\alpha\,m_P\;,
$$
which is Planckian. Thus, we expect \mbox{$r\sim 0.01$} from the Lyth bound.} 
\begin{equation}
\varphi_{*}=-\sqrt{\alpha}\ln\left[\frac{\kappa}{\alpha}\left(N_{*}+\sqrt{\frac{\alpha}{2}}\right)\right]\mpl\,,\label{eq:phi_star}
\end{equation}
where $N_{*}$ is the number of e-folds of inflation since the pivot
scale exits the horizon. This gives us an idea of the value of the
slow-roll parameters as a function of $N_{*}$ 
\begin{align}
\epsilon_{*} & = \frac{\alpha/2}{\left(N_{*}+\sqrt{\alpha/2}\right)^{2}}\simeq\frac{\alpha}{2N_{*}^{2}}\,,\label{eps-star}\\
\eta_{*} & = \frac{\alpha-N_{*}-\sqrt{\alpha/2}}{\left(N_{*}+\sqrt{\alpha/2}\right)^{2}}\simeq\frac{\alpha}{N_{*}^{2}}-\frac{1}{N_{*}}
\end{align}
which it is nice to note are independent of $\kappa$. The spectral
index and tensor to scalar ratio are hence also independent of $\kappa$
and given by 
\begin{eqnarray}
n_{\mathrm{s}} & = & 1+2\eta_{*}-6\epsilon_{*}\nonumber \\
 & \simeq & 1-\frac{\alpha}{N_{*}^{2}}-\frac{2}{N_{*}}\,,\label{spectral}\\
r & = & 16\epsilon_{*}\simeq\frac{8\alpha}{N_{*}^{2}}\,.\label{ratio}
\end{eqnarray}
As we would expect for a plateau inflation model, they match the Planck
results \citep{Ade:2015lrj} exceptionally well for a range of parameter
values. The limiting observational constraint is $n_{s}=0.968\pm0.006$
($2\sigma$ result) and the upper bound on the tensor to scalar ratio
$r<r_{\mathrm{bound}}$, where $r_{\mathrm{bound}}=0.06$ (at $2\sigma$
confidence level).

Using eqs.~\eqref{spectral} and \eqref{ratio} we can compute the
upper bound on $N_{*}$. Assuming $N_{*}>31$, we find 
\begin{align}
N_{*}< & \frac{16}{8\left(1-n_{\mathrm{s}}\right)-r_{\mathrm{bound}}}\,.
\end{align}
Moreover, using Eq.~\eqref{spectral} we can relate $\alpha$ to
the number of e-folds $N_{*}$: 
\begin{equation}
\alpha=\left(1-n_{\mathrm{s}}\right)N_{*}^{2}-2N_{*}\,.\label{aN}
\end{equation}

Constraining the maximum possible value\footnote{We can estimate this as follows. The value of $N_{*}$ is increased
by considering that, after inflation there is an ``stiff'' epoch
when the barotropic parameter of the Universe is %
\mbox{%
$\frac{1}{3}<w\leq1$%
}. The upper bound ensures that the speed of sound is not superluminal.
The larger $w$ is and the longer this stiff period lasts, the more
$N_{*}$ becomes. Thus, we can take %
\mbox{%
$w=1$%
} as in kination, and we can assume that kination starts immediately
after the end of inflation and until reheating. Then we have 
\[
N_{*}=57+\frac{1}{3}\ln\left(\frac{V_{\mathrm{end}}^{1/4}}{T_{\mathrm{reh}}}\right)\,,
\]
 where $T_{{\rm reh}}$ is the reheating temperature. The observational
bound on the inflation scale is %
\mbox{%
$V_{{\rm end}}^{1/4}\lesssim10^{16}\,$GeV%
}. Saturating this bound, while considering the lowest $T_{{\rm reh}}$
possible (%
\mbox{%
$T_{{\rm reh}}\gtrsim10\,$MeV%
}, to avoid spoiling Big Bang Nucleosynthesis) we find %
\mbox{%
$N_{*}^{{\rm max}}\simeq70$%
}.} of $N_{*}<70$ and $\alpha\ge1$ we find the allowed range for $N_{*}$
and $\alpha$ which is compatible with CMB observations to be 
\begin{align}
53\le N_{*} & <70\,,\label{N-bound}\\
1\le\alpha & <37\,.\label{a-bound}
\end{align}
Note that the allowed parameter range above depends sensitively on
the value of $n_{\mathrm{s}}$. The above region is the maximum region
within the $2\sigma$ range of $n_{\mathrm{s}}$. For comparison, fixing
$n_{\mathrm{s}}$ at the best fit value $n_{s}=0.968$ gives %
\mbox{%
$63\le N_{*}<70$%
} and $1\le\alpha<15.8$.

\begin{figure}
\begin{centering}
\includegraphics[scale=0.4]{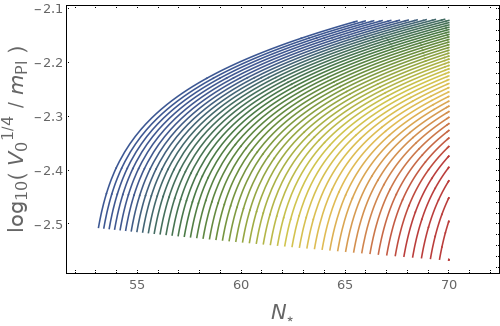} ~ \includegraphics[scale=0.4]{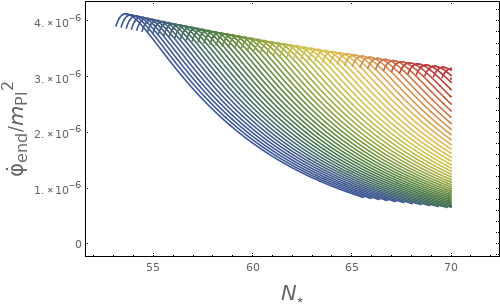} 
\par\end{centering}
\caption{\label{fig:VN}The allowed range of $V_{0}$ values (upper) and the
field velocity at the end of inflation $\dot{\varphi}_{\mathrm{end}}$
(lower) as a function of $N_{*}$. Each curve represents a fixed value
of the scalar spectral index $n_{\mathrm{s}}$. The blue leftmost
curve is for the $2\sigma$ lower bound $n_{s}=0.962$ and the red
rightmost curve is for $n_{\mathrm{s}}=0.971$. We used the central
value $\ln\left(10^{10}A_{\mathrm{s}}\right)=3.094$ of the spectrum
normalisation in Eq.~\eqref{As-val}.}
\end{figure}

The normalisation of the power spectrum further constrains the model.
The amplitude of the scalar spectrum at the pivot scale is given by
\begin{equation}
A_{\mathrm{s}}=\frac{1}{24\pi^{2}\mpl^{4}}\frac{V}{\epsilon}\,.\label{As}
\end{equation}
Plugging eqs.~\eqref{V-infl}, \eqref{eq:phi_star} and \eqref{eps-star}
into the expression above we find 
\begin{equation}
\frac{V_{0}}{\mpl^{4}}=\frac{12\pi^{2}}{\mathrm{e}^{2}}\left[\left(1-n_{\mathrm{s}}\right)-\frac{2}{N_{*}}\right]\mathrm{e}^{\left(1-n_{\mathrm{s}}\right)N_{*}}A_{\mathrm{s}}\,,\label{V0As}
\end{equation}
where we used Eq.~\eqref{aN}. This expression relates the energy
scale of inflation to the number of e-folds $N_{*}$. The Planck collaboration
reports the value of $A_{\mathrm{s}}$ as \citep{Ade:2015lrj}
\begin{equation}
\ln\left(10^{10}A_{\mathrm{s}}\right)=3.094\pm0.034\label{As-val}
\end{equation}
(at the $1\sigma$ confidence level). On the upper panel of Fig.~\ref{fig:VN}
we display the allowed range of $V_{0}$ for the best fit value of
$A_{\mathrm{s}}$. 

Using $A_{\mathrm{s}}$ in Eq.~\eqref{As-val} and plugging Eq.~\eqref{V0As}
into \eqref{dphi-end} we can find $\dot{\varphi}_{\mathrm{end}}$.
For the best fit value of $A_{\mathrm{s}}$ the allowed range of $\dot{\varphi}_{\mathrm{end}}$
is show on the lower panel of Fig.~\ref{fig:VN}. We see that the
range is 
\begin{equation}
0.6<\frac{\dot{\varphi}_{\mathrm{end}}}{\mpl^{2}}\times10^{6}<4.2\,.\label{dphiend-b}
\end{equation}

The $\kappa$ parameter in Eq.~\eqref{L-org} determines the ratio
of inflation energy density to the vacuum energy density.%
\footnote{As we have shown in footnote~3, for the non-canonical field during
  inflation we have that $\phi\sim\sqrt\alpha m_P/\kappa$. Thus, for the mass
  of the $\chi$-field during inflation from Eq.~\eqref{L-org} we find
  $m_\chi\sim(g\sqrt\alpha/\kappa)m_P$. Because $\alpha,g\sim 1$ and
  $\kappa\sim 100$ we have that the mass of the $\chi$-field during inflation is
  $m_\chi\sim 10^{-2}m_P$, i.e. comparable to the scale of grand unification
  and bigger that the Hubble scale during inflation. This means that the
  $\chi$-field is heavy during inflation, as we have assumed. In
  contrast, the inflaton field is light because $\eta$, calculated in 
  Eq.~\eqref{eta}, is much smaller than unity and this is why the value of the
  spectral index is close to unity.}
Plugging $\phi=\phi_{\esp}$ and $\chi=f$ into that expression gives 
\begin{equation}
\kappa=\frac{\mpl}{\phi_{\esp}}\ln\frac{V_{0}}{V_{\vac}}\,,\label{k-Vs}
\end{equation}
where $V_{\vac}=V\left(\phi=\phi_{\esp},\chi=f\right)\simeq10^{-12}\:\mathrm{eV}^{4}\simeq10^{-120}\mpl^{4}$
is the vacuum energy density. Taking $V_{0}^{1/4}<10^{-2}\mpl$ (see
Fig.~\ref{fig:VN}) we find 
\begin{equation}
\kappa\frac{\phi_{\esp}}{\mpl}<261\,.\label{kappa-b}
\end{equation}

\section{Kination\label{subsec:Kination}}

Once inflation ends, almost immediately the period of kination sets
in.\footnote{We have confirmed this using numerical simulations.}
During kination the field is oblivious of the potential and the Klein-Gordon
(KG) equation takes the form 
\begin{equation}
\ddot{\varphi}+3H\dot{\varphi}\simeq0\,,\label{eq:kination_KG}
\end{equation}
and the Friedmann equation is
\begin{equation}
3\mpl^{2}H^{2}=\frac{\dot{\varphi}^{2}}{2}\,.\label{eq:kination_fried}
\end{equation}
Substituting Eq. \eqref{eq:kination_fried} into Eq. \eqref{eq:kination_KG}
gives
\begin{equation}
\ddot{\varphi}+\sqrt{\frac{3}{2}}\frac{\dot{\varphi}^{2}}{\mpl}=0\,.
\end{equation}
Integrating the above equation, we find the solutions 
\begin{align}
\varphi & =\varphi_{0}+\sqrt{\frac{2}{3}}\mpl\ln\left[1+\sqrt{\frac{3}{2}}\frac{\dot{\varphi}_{0}}{\mpl}\left(t-t_{0}\right)\right]\,,\label{phi_general}\\
\dot{\varphi} & =\dot{\varphi}_{0}\,\mathrm{exp}\left[-\sqrt{\frac{3}{2}}\frac{\left(\varphi-\varphi_{0}\right)}{\mpl}\right]\,.\label{phidot_general}
\end{align}
where the subscript `0' refers to the initial value in the integration.
In our case $\varphi_{0}\simeq\varphi_{\mathrm{end}}$ and $\dot{\varphi}_{0}\simeq\dot{\varphi}_{\mathrm{end}}$
in eqs.~\eqref{phi-end} and \eqref{dphi-end} respectively, meaning
the above equations become 
\begin{align}
\varphi_{\mathrm{kin}} & =\mpl\sqrt{\alpha}\,\mathrm{\ln}\,\frac{\sqrt{2\alpha}}{\kappa}+\sqrt{\frac{2}{3}}\mpl\ln\left[1+\frac{\sqrt{V_{\mathrm{end}}}}{\mpl}\left(t-t_{0}\right)\right]\,,\label{kin_phi_2}\\
\dot{\varphi}_{\mathrm{kin}} & =\sqrt{\frac{2}{3}V_{\mathrm{end}}}\left(\frac{\sqrt{2\alpha}}{\kappa}\right)^{\sqrt{\frac{3\alpha}{2}}}\mathrm{e}^{-\sqrt{\frac{3}{2}}\frac{\varphi}{\mpl}}\,.\label{kin_phidot_2}
\end{align}
Taking $\varphi_{0}\simeq\varphi_{\mathrm{end}}$ and $\dot{\varphi}_{0}\simeq\dot{\varphi}_{\mathrm{end}}$
presumes an immediate transition from inflation to kination, when $\epsilon=1$.

During kination, the kinetic energy density of the inflaton scales
as $\dot{\varphi}^{2}/2\propto a^{-6}$, where $a$ is the scale factor.
Therefore the false vacuum energy density $\lambda f^{4}/4$ might
come to dominate at some later times. For the sake of simplicity we
restrict the model to the parameter range where this never happens
until $\varphi$ reaches ESP.

To find this regime we can compute $\dot{\varphi}_{\esp}$ using Eq.~\eqref{kin_phidot_2}.
Plugging eqs.~\eqref{phivarphi} and \eqref{Vend} into the latter
we obtain 
\begin{align}
\dot{\varphi}_{\esp} & =\frac{\sqrt{2V_{0}\left(\alpha\right)/3}}{\mathrm{e}^{\sqrt{\alpha/2}}}\left(\frac{\sqrt{2\alpha}\,m_{{\rm Pl}}}{\kappa\left(\alpha\right)\phi_{\esp}}\right)^{\sqrt{\frac{3\alpha}{2}}}\,,\label{dphi-esp}
\end{align}
where the argument in $V_{0}\left(\alpha\right)$ and $\kappa\left(\alpha\right)$
is to remind us that observations constrain $\kappa$ and $V_{0}$
to be functions of $\alpha$ via eqs.~\eqref{k-Vs}, \eqref{V0As}
and \eqref{aN}.

Using Eq.~\eqref{dphi-esp}, we plot $\dot{\varphi}_{\esp}/2$ as
a function of $\alpha$ in Fig.~\ref{fig:kindom} along with the
constant value of $\lambda f^{4}/4$. As we can easily see the universe
is in the regime of kination at the ESP if 
\begin{equation}
\frac{\lambda f^{4}}{2\dot{\varphi}_{\esp}^{2}}<1\:\Rightarrow\:\alpha<10\,.\label{eng-rat}
\end{equation}
The precise value of course depends on $f$ and (only mildly) on $n_{\mathrm{s}}$,
but we adopt this bound as our reference value.

\begin{figure}
\begin{centering}
\par\end{centering}
\caption{\label{fig:kindom}The inflaton kinetic energy density at ESP (blue
curve) and the false vacuum energy density (green line) as a function
of $\alpha$. In this plot we used the best fit value of $n_{\mathrm{s}}$
(within $2\sigma$ range the result depends mildly on the precise
value of $n_{\mathrm{s}}$) and we took $\lambda=1$ and $f=10^{12}\:\mathrm{GeV}$.}
\end{figure}

\section{Inflaton trapping at the ESP\label{subsec:prm}}

During the period of kination, the inflaton follows the equation of
motion \eqref{eq:kination_KG}. Once it approaches $\phi\rightarrow\phi_{\esp}$,
the tachyonic and resonant excitations of the $\chi$ field produce
large numbers of particles. These particles backreact onto the motion
of the inflaton, creating an effective linear potential for the latter.
If the production of $\chi$ particles is efficient, then the inflaton's fast
rolling is halted by trapping it at the ESP.

As we are going to see, the trapping can be very abrupt. Therefore,
it is a good approximation to neglect the expansion of the universe.
Moreover, as we will show \textit{a posteriori}, $\varphi$ oscillates
around the ESP with an amplitude much smaller than $\mpl$,
$\left|\varphi-\varphi_{\esp}\right|\ll\mpl$.
Therefore, in considerations of the trapping process, it is enough
to study the Lagrangian up to the first order in $\left|\varphi-\varphi_{\esp}\right|/\mpl$.
In this case, we find from Eq.~\eqref{eq:lagrangian_varphi} 
\begin{equation}
V\simeq V_{\vac}\left(\frac{V_{\vac}}{V_{0}}\right)^{\frac{\varphi-\varphi_{\mathrm{ESP}}}{\sqrt{\alpha}\mpl}}+\frac{1}{2}\gamma^{2}\left(\varphi-\varphi_{\mathrm{ESP}}\right)^{2}\chi^{2}+V\left(\chi\right)\,,\label{Vesp}
\end{equation}
where 
\begin{equation}
\gamma^{2}\equiv\frac{g^{2}\phi_{\esp}^{2}}{\alpha\mpl^{2}}\label{gamma-def}
\end{equation}
is the effective quartic coupling constant close to the ESP. The first
term in Eq.~\eqref{Vesp} sets the scale of the vacuum energy. It
is much smaller than $\lambda f^{4}/4$ and can be neglected during
the inflaton trapping phase.

With the above assumptions in mind we can write the equation of motion
for the inflaton as 
\begin{equation}
\ddot{\varphi}+\gamma^{2}\left\langle \chi^{2}\right\rangle \left(\varphi-\varphi_{\mathrm{ESP}}\right)\simeq0\,.\label{EoM-infl}
\end{equation}
Initially the expectation value of the $\chi$ field is zero, which
makes $\varphi$ massless at the classical level. However, quantum
corrections due to the interaction term generate an effective mass
for the $\varphi$ field. At the first order such corrections can
be accounted for by using the Hartree approximation, which has been
employed in the expression above.

The trapping of the inflaton $\varphi$ by the resonant production
of $\chi$ particles has been studied in great detail in
Ref.~\citep{Kofman:2004yc}.
In that work, the primary source of particle production is the parametric
resonance. In our case, $\omega_{k}^{2}$ (defined below) is not positive
definite, therefore the $\chi$ field is also excited by the tachyonic
instability close to the ESP. Moreover, in our model the 
quartic self-interaction of the $\chi$ field
can affect the particle production too. To investigate these processes
we start by writing the equation of motion for the mode functions
of the $\chi$ field: 
\begin{equation}
\ddot{\chi}_{k}+\omega_{k}^{2}\chi_{k}=0\,,\label{EoM-chi-apprx}
\end{equation}
where 
\begin{equation}
\omega_{k}^{2}=k^{2}+\gamma^{2}\left(\varphi-\varphi_{\mathrm{ESP}}\right)^{2}+\lambda\left(3\left\langle \chi^{2}\right\rangle -f^{2}\right)\,.\label{freq}
\end{equation}
Initially $\left\langle \chi^{2}\right\rangle $ is negligible and
we can clearly see in the above equation that $\chi_{k}$ modes with
$k<k_{\c}$ acquire an effective imaginary mass as $\varphi\rightarrow\varphi_{\esp}$,
where $k_{\c}$ is the critical wavenumber 
\begin{equation}
k_{\c}\equiv\sqrt{\lambda}f\,.
\end{equation}
These modes are unstable and start growing due to the tachyonic instability.
The process is similar to the one analysed in Ref.~\citep{Dufaux:2006ee}.
To compute the production of $\chi$ particles note, first, that in
a narrow window 
\begin{equation}
\left|\varphi-\varphi_{\mathrm{ESP}}\right|\le\frac{\sqrt{\lambda}f}{\gamma}
\end{equation}
we can approximate the evolution of $\varphi$ linearly (see Eq.~\eqref{phi_general})
\begin{equation}
\varphi-\varphi_{\mathrm{ESP}}\simeq\dot{\varphi}_{\esp}\tau\,,\label{phi-lin}
\end{equation}
where $\tau\equiv t-t_{\esp}$ and $t_{\esp}$ is defined as $\varphi\left(t_{\esp}\right)\equiv\varphi_{\esp}$
and $\dot{\varphi}_{\esp}$ in the above equation is the field velocity
at the ESP from Eq.~\eqref{dphi-esp}.

Using Eq.~\eqref{phi-lin} we find that after the first crossing
of the ESP the occupation number of $\chi_{k}$ mode is given by \citep{Dufaux:2006ee}
\begin{equation}
  n_{k}=\exp 
  \intop_{\tau_{-}}^{\tau_{+}}2
  \sqrt{-\omega^{2}\left(\tau'\right)}\,\mathrm{d}\tau'\label{Xk-def}
\end{equation}
and $\omega^{2}\left(\tau_{\pm}\right)=0$, that is 
\begin{equation}
\tau_{\pm}=\pm\sqrt{\frac{\lambda f^{2}-k^{2}}{\gamma^{2}\dot{\varphi}_{\esp}^{2}}}\,.
\end{equation}
It is easy to integrate Eq.~\eqref{Xk-def}, which gives 
\begin{equation}
n_{k}=\mathrm{e}^{\pi\frac{\lambda f^{2}-k^{2}}{\gamma\dot{\varphi}_{\esp}}}\,.
\end{equation}

Integrating over all $k$ wavenumbers, we obtain the total occupation
number of newly produced particles after the first passage of the ESP
\begin{align}
  n_{\mathrm{\chi1}} & \simeq\frac{1}{2\pi^{2}}\intop_{0}^{\infty}
  n_{k}k^{2}\mathrm{d}k\\
 & \simeq\frac{\left(\gamma\dot{\varphi}_{\esp}\right)^{3/2}}{\left(2\pi\right)^{3}}\left[-2s^{1/4}+\mathrm{e}^{\pi s^{1/2}}\mathrm{Erf}\left(\sqrt{\pi}s^{1/4}\right)\right]\,,
\end{align}
where $\mathrm{Erf}\left(x\right)\equiv2/\sqrt{\pi}\intop_{0}^{z}\mathrm{e}^{-t^{2}}\mathrm{d}t$
is the error function and 
\begin{equation}
s\equiv\left(\frac{\lambda f^{2}}{\gamma\dot{\varphi}_{\esp}}+\frac{2}{3^{3/2}}\right)^{2}\,.
\end{equation}
The first term in the parenthesis is positive. Therefore,
$s^{1/4}\ge\sqrt{2}/3^{3/4}$
and $\mathrm{Erf}\left(\sqrt{\pi}s^{1/4}\right)\simeq1$. Moreover,
$\exp\{\pi\sqrt{s}\}>s^{1/4}$ and we can write 
\begin{equation}
n_{\chi1}\simeq\frac{\left(\gamma\dot{\varphi}_{\esp}\right)^{3/2}}{\left(2\pi\right)^{3}}\mathrm{e}^{\pi\left(\frac{\lambda f^{2}}{\gamma\dot{\varphi}_{\esp}}+\frac{2}{3^{3/2}}\right)}\,.\label{n1}
\end{equation}

Similarly we can compute the dispersion $\left\langle \chi^{2}\right\rangle $
\citep{Kofman:1997yn} 
\begin{align}
\left\langle \chi^{2}\right\rangle  & =\frac{1}{2\pi^{2}}\int\frac{n_{k}k^{2}}{\omega_{k}}\mathrm{d}k\\
 & \simeq\frac{n_{\chi}}{\gamma\left|\varphi-\varphi_{\mathrm{ESP}}\right|}\,,\label{chi2}
\end{align}
where we used the fact that $\gamma\left(\varphi-\varphi_{\mathrm{ESP}}\right)>k_{\c}$
in the non-tachyonic regime.

Plugging Eq.~\eqref{chi2} into \eqref{EoM-infl}, we can write the
inflaton equation of motion as 
\begin{equation}
\ddot{\varphi}+\gamma n_{\chi}\mathrm{sign}\left(\varphi-\varphi_{\mathrm{ESP}}\right)\simeq0\,,\label{EoM-trap}
\end{equation}
where $\mathrm{sign}\left(\varphi-\varphi_{\mathrm{ESP}}\right)$
is the signature of $\varphi-\varphi_{\mathrm{ESP}}$. This equation
describes oscillations in a linear potential: at the time 
\begin{equation}
\tau_{1}\equiv\frac{\dot{\varphi}_{\esp}}{\gamma n_{\chi1}}\label{tau1}
\end{equation}
the field $\varphi$ reaches the value $\varphi_{1}=\varphi_{\esp}+\Phi_{1}$
and rolls back toward $\varphi_{\esp}$, where 
\begin{equation}
\Phi_{1}\simeq\frac{1}{2}\frac{\dot{\varphi}_{\esp}^{2}}{\gamma n_{\chi1}}\label{phi1}
\end{equation}
is the amplitude of the first oscillation.

\begin{figure}
\begin{centering}
\includegraphics[scale=0.5]{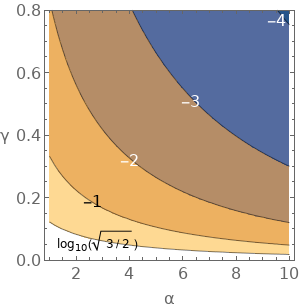}
\par\end{centering}
\caption{\label{fig:sdn}A contour plot of $\log_{10}\left(\Phi_{1}/\protect\mpl\right)$
values as a function of $\alpha$ and $\gamma$ (see Eq.~\ref{Phi1-expl}).
In this plot we used $\lambda f^{2}/\gamma\dot{\varphi}_{\protect\esp}\ll2/3^{3/2}$
and the best fit value $n_{\mathrm{s}}=0.968$. The plot changes insignificantly
within the allowed $2\sigma$ range of $n_{\mathrm{s}}$. The white
region is where the condition in Eq.~\eqref{cond-sml} is violated,
i.e. where the expansion of the Universe cannot be neglected.}
\end{figure}

Up to now we have assumed that the expansion of the universe does
not affect the trapping process. This is justified if the timescale
$\tau_{1}\simeq2\left(\varphi_{1}-\varphi_{\esp}\right)/\dot{\varphi}_{\esp}$
is much shorter than the Hubble time $H_{\esp}^{-1}$ at the ESP; that
is $H_{\esp}\ll\dot{\varphi}_{\esp}/2\Phi_{1}$. Using $H_{\esp}\simeq\dot{\varphi}_{\esp}/\sqrt{6}\mpl$,
in agreement with our assumption $\alpha<10$ (see Fig.~\ref{fig:kindom}),
we find that this is the case when 
\begin{equation}
\Phi_{1}\ll\sqrt{\frac{3}{2}}\mpl\,.\label{cond-sml}
\end{equation}
Plugging eqs.~\eqref{dphi-esp} and \eqref{n1} into Eq.~\eqref{phi1},
we find 
\begin{equation}
\Phi_{1}\simeq\frac{4\pi^{3}}{\gamma^{5/2}}\frac{\left[\frac{2}{3}V_{0}\left(\alpha\right)\right]^{1/4}}{\mathrm{e}^{\pi\left(\frac{\lambda f^{2}}{\gamma\dot{\varphi}_{\esp}}+\frac{2}{3^{3/2}}\right)}}\left(\frac{\sqrt{2\alpha}}{\mathrm{e}^{1/\sqrt{3}}}\frac{\mpl}{\kappa\left(\alpha\right)\phi_{\esp}}\right)^{\frac{1}{2}\sqrt{\frac{3\alpha}{2}}}\,.\label{Phi1-expl}
\end{equation}
To find the range where Eq.~\eqref{cond-sml} is satisfied we plot
$\log_{10}\Phi_{1}$ as a function of $\alpha$ and $\gamma$ in Fig.~\ref{fig:sdn}.
We can see that Eq.~\eqref{cond-sml} can be easily satisfied for
reasonable values of $\gamma$.

Each time $\varphi$ crosses the ESP, the $\chi$ field experiences a burst
of particle production. New particles strengthen the backreaction
onto the motion of $\varphi$, causing an exponential decay of the
oscillation amplitude. Via the quartic self-interaction term these
particles also generate a contribution to the effective mass of the
$\chi$ field. At the level of the Hartree approximation these interactions
are taken care of by the $3\lambda\left\langle \chi^{2}\right\rangle $
term in Eq.~\eqref{freq}. The tachyonic particle production is effective
as long as the $3\lambda\left\langle \chi^{2}\right\rangle $ term
is smaller than $\lambda f^{2}$.

To compute the occupation number at the end of the tachyonic instability
we first find the minimum of $\omega_{k=0}^{2}$ in Eq.~\eqref{freq}
which is located at $\left[\gamma\left(\varphi_{\mathrm{min}}-\varphi_{\mathrm{ESP}}\right)\right]^{3}=3\lambda n_{\chi}/2$,
where we used Eq.~\eqref{chi2}. Plugging this back into the expression
of $\omega_{k=0}^{2}$ and equating it to zero, we find that the tachyonic
resonance stops at 
\begin{equation}
n_{\chi,\mathrm{tach}}\simeq\frac{2}{3^{5/2}}\sqrt{\lambda}f^{3}\,\cdot
\end{equation}
At this moment the oscillation amplitude is given by 
\begin{equation}
\Phi_{\mathrm{tach}}\simeq\frac{3^{5/2}}{4}\frac{\dot{\varphi}_{\esp}^{2}}{\gamma\sqrt{\lambda}f^{3}}\,.
\end{equation}
We can compare $n_{\chi,\mathrm{tach}}$ with the number density produced
after the first oscillation in Eq.~\eqref{n1} 
\begin{equation}
\frac{n_{\chi1}}{n_{\chi,\mathrm{tach}}}\simeq\frac{3^{5/2}}{16\pi^{3}}\lambda\left(\frac{\gamma\dot{\varphi}_{\esp}}{\lambda f^{2}}\right)^{3/2}\mathrm{e}^{\pi\left(\frac{\lambda f^{2}}{\gamma\dot{\varphi}_{\esp}}+\frac{2}{3^{3/2}}\right)}\,.
\end{equation}
For $\gamma\dot{\varphi}_{\esp}/\lambda f^{2}=2\pi/3$ the right-hand side 
of the above expression is minimal and given by 
\begin{equation}
\left.\frac{n_{\chi1}}{n_{\chi,\mathrm{tach}}}\right|_{\mathrm{min}}\sim\lambda\lesssim1\,.
\end{equation}
Thus, depending on the magnitude of $\gamma\dot{\varphi}_{\esp}/\lambda f^{2}$,
it might take several passages through the ESP before the tachyonic resonance
is terminated.

The end of the tachyonic resonance does not necessarily imply the
end of particle production though. After $\omega_{k}^{2}$ becomes
positive definite for all values of $k$, particles may still be produced
by the parametric resonance. Such a production continues as long as
the adiabaticity condition 
\begin{equation}
\frac{\left|\dot{\omega}_{k}\right|}{\omega_{k}^{2}}<1\label{adb-cond}
\end{equation}
is broken within some range of $\varphi$ values $\Delta\varphi_{\mathrm{nad}}$,
that is $\left|\dot{\omega}_{k}\right|/\omega_{k}^{2}\ge1$ for $\varphi-\varphi_{\esp}\in\left[-\Delta\varphi_{\mathrm{nad}},\Delta\varphi_{\mathrm{nad}}\right]/2$.
As $\left\langle \chi^{2}\right\rangle $ continues to grow with each
burst of particle production, it will eventually shut down the parametric
resonance too.

The shut-down is caused by one of the two effects, whichever happens
first: either the oscillation amplitude $\Phi$ becomes smaller than
$\Delta\varphi_{\mathrm{nad}}$ \citep{Kofman:2004yc} (Case 1) or
quartic self-interactions of the $\chi$ field render it too heavy
to be excited (Case 2). The choice between the two cases is determined
by the magnitude of the ratio $\lambda/\gamma^{2}$.

Let us consider these two possibilities in turn. To do that we can
safely employ Eq.~\eqref{phi-lin} within the narrow non-adiabaticity
window. Hence we can write 
\begin{equation}
\frac{\dot{\omega}_{k}}{\omega_{k}^{2}}\simeq\frac{\gamma^{2}\left(\varphi-\varphi_{\mathrm{ESP}}\right)\dot{\varphi}_{\esp}}{\left[k^{2}+\gamma^{2}\left(\varphi-\varphi_{\mathrm{ESP}}\right)^{2}+\lambda\left(3\left\langle \chi^{2}\right\rangle -f^{2}\right)\right]^{3/2}}\,,\label{rez}
\end{equation}
where in the adiabatic regime $\left\langle \chi^{2}\right\rangle $
is given by the expression in Eq.~\eqref{chi2}.

In Case 1 the resonance stops before the $\lambda\left(3\left\langle \chi^{2}\right\rangle -f^{2}\right)$
term in Eq.~\eqref{rez} becomes important and we neglect it. This
gives $\Delta\varphi_{\mathrm{nad}}\simeq\left(\dot{\varphi}_{\esp}/\gamma\right)^{1/2}$.
Once the oscillation amplitude $\Phi=\dot{\varphi}_{\esp}^{2}/2\gamma n_{\chi}$
drops below this value, particles no longer grow via the process
of parametric resonance. Equating $\Delta\varphi_{\mathrm{nad}}=\Phi$
we find 
\begin{equation}
\Phi_{\mathrm{fin}1}\simeq\left(\frac{\dot{\varphi}_{\esp}}{\gamma}\right)^{1/2}\label{Phifin1}
\end{equation}
and 
\begin{equation}
n_{\chi,\mathrm{fin1}}\simeq\frac{1}{2}\frac{\dot{\varphi}_{\esp}^{3/2}}{\gamma^{1/2}}\,,\label{nfin1}
\end{equation}
where the subscript `$\mathrm{fin}1$' signifies the case where the
resonance stops because the inflaton oscillation amplitude drops bellow
$\Delta\varphi_{\mathrm{nad}}$. We can compute the energy density
\mbox{%
$\rho_{\chi\mathrm{fin}1}\simeq\gamma\Phi_{\mathrm{fin}1}n_{\chi,\mathrm{fin1}}$%
} in the $\chi$ particles at that moment 
\begin{equation}
\rho_{\chi\mathrm{fin}1}\simeq\frac{1}{2}\dot{\varphi}_{\esp}^{2}\,,\label{rhofin1}
\end{equation}
which is about the same as the initial kinetic energy density of the
inflaton when it first crosses the ESP.

In Case 1 we could neglect $\chi$ field self-interactions in Eq.~\eqref{rez}.
This rendered $\Delta\varphi_{\mathrm{nad}}\simeq\mathrm{constant}$.
On the other hand, if $\lambda/\gamma^{2}$ is large, which corresponds
to Case 2, such self-interactions cannot be neglected and the $\left\langle \chi^{2}\right\rangle $
term in Eq.~\eqref{rez} becomes significant. As the importance of
this term grows, the non-adiabaticity region $\Delta\varphi_{\mathrm{nad}}$
shrinks to zero eventually halting the resonance.

To estimate the end of the resonance we find the moment when the maximum
value of the ratio in Eq.~\eqref{rez} becomes smaller than one.
For the $k=0$ mode the maximum value of this ratio is approximately
\begin{equation}
\left.\frac{\dot{\omega}_{k=0}}{\omega_{k=0}^{2}}\right|_{\mathrm{max}}\simeq\frac{\gamma\dot{\varphi}_{\esp}}{6\left(\lambda n_{\chi}\right)^{2/3}}\,,
\end{equation}
where we used $3\left\langle \chi^{2}\right\rangle >f^{2}$ and Eq.~\eqref{chi2}.
The resonance becomes inefficient once this value falls bellow unity.
Hence, we can consider the particle production to be over when 
\begin{equation}
n_{\chi,\mathrm{fin2}}\simeq\frac{\left(\gamma\dot{\varphi}_{\esp}/6\right)^{3/2}}{\lambda}\,.\label{nfin2}
\end{equation}
Plugging this value into Eq.~\eqref{EoM-trap} we find that the inflaton
oscillation amplitude at this point is 
\begin{equation}
\Phi_{\mathrm{fin2}}\simeq7\frac{\lambda}{\gamma^{2}}\left(\frac{\dot{\varphi}_{\esp}}{\gamma}\right)^{1/2}\,.\label{Phifin2}
\end{equation}

Comparing Eq.~\eqref{nfin1} with \eqref{nfin2} and Eq.~\eqref{Phifin1}
with \eqref{Phifin2} we see that the first mechanism is responsible
for the end of the resonance if 
\begin{equation}
  \frac{\lambda}{\gamma^{2}}<\frac{1}{7}\,.
  \label{bound}
\end{equation}
In this case the energy density of $\chi$ particles is comparable
to the inflaton's initial kinetic energy (see Eq.~\eqref{rhofin1}).
In the opposite regime, the strong quartic self-interaction $\lambda\chi^{4}$
shuts down the resonance much earlier, leaving a larger fraction of
the energy budget in the inflaton sector. Moreover, the inflaton oscillation
amplitude is larger too. In summary, a stronger $\chi$ field self-interaction
results in less efficient inflaton trapping.

\section{Reheating}

As we have shown, after crossing the ESP the total kinetic
density of the inflaton decays into radiation through resonant production
of $\chi$-particles.\footnote{The case when the quartic self-interaction
  of the $\chi$-particles stops their resonant production early (Case~2,
  discussed above) introduces the extra complication of the perturbative decay
  of the oscillating inflaton condensate. For simplicity, we consider only
  Case~1, which amounts to satisfying the bound in Eq.~\eqref{bound}.}
Thus we expect 
\begin{equation}
\frac{1}{2}\dot{\varphi}_{{\rm ESP}}^{2}\simeq\frac{\pi^{2}}{30}g_{*}T_{{\rm reh}}^{4}\,,\label{dphi-Treh}
\end{equation}
where %
\mbox{%
$g_{*}={\cal O}(100)$%
} is the effective relativistic degrees of freedom. Using the above
and Eq.~\eqref{dphi-esp} we obtain 
\begin{equation}
T_{{\rm reh}}=\left[\frac{10}{\pi^{2}g_{*}}V_{{\rm end}}\left(\frac{\sqrt{2\alpha}}{\kappa}\frac{m_{{\rm Pl}}}{\phi_{{\rm ESP}}}\right)^{\sqrt{6\alpha}}\right]^{1/4},\label{Treh}
\end{equation}
where we used %
\mbox{%
$V_{{\rm end}}=V_{0}e^{-\sqrt{2\alpha}}$%
} according to Eq.~\eqref{Vend}. To get a feeling about the magnitude
of $T_{{\rm reh}}$ we take %
\mbox{%
$\phi_{{\rm ESP}}=m_{{\rm Pl}}$%
} and %
\mbox{%
$V_{{\rm end}}^{1/4}\sim V_{0}^{1/4}\sim10^{-2}\,m_{{\rm Pl}}$%
}, in which case we have 
\begin{equation}
T_{{\rm reh}}\sim10^{-3}m_{{\rm Pl}}\left(\frac{\sqrt{2\alpha}}{\kappa}\right)^{\sqrt{3\alpha/8}}\,.
\end{equation}
As we show below %
\mbox{%
$\kappa={\cal O}(100)$%
} in order to have successful dark energy. Then %
\mbox{%
$\alpha\in[1.5,10]$%
} gives %
\mbox{%
$T_{{\rm reh}}\sim10^{11-13}\,$GeV%
}. This means that, if $T_{\rm reh}$  is large, thermal corrections might
restore the Peccei-Quinn
symmetry, unless $\kappa$ is rather large, approaching the bound
in Eq.~\eqref{kappa-b}. Actually, in this case, the Peccei-Quinn
symmetry is thermally broken later on, after the onset of radiation
domination. In contrast, if the reheating temperature is not very
large, the Peccei-Quinn symmetry is broken once the inflaton rolls
towards the VEV regardless of thermal corrections. In both cases,
the axion does not exist during inflation and so there is no problem
with axion isocurvature perturbations.

Knowing the reheating temperature in Eq.~\eqref{Treh}, we can calculate
the number of e-folds of inflation from the recognisable equation
\begin{multline}
N_{*}=61.2+\frac{1}{3(1+w)}\ln\left(\frac{g_{*}\pi^{2}}{60}\right)+\\
\frac{(3w-1)}{3(1+w)}\ln\left(\frac{V_{\mathrm{end}}^{1/4}}{T_{\mathrm{reh}}}\right)+\ln\left(\frac{V_{\mathrm{end}}^{1/4}}{\mpl}\right)\,.\label{eq:efolds_final}
\end{multline}
Presuming the kinetically dominated inflaton field is still the dominant
component of the energy density until the produced radiation dominates,
we have $w=1$ and the above becomes 
\begin{equation}
N_{*}=61.7+\frac{1}{3}\ln\left(\frac{V_{\mathrm{end}}^{1/4}}{T_{\mathrm{reh}}}\right)+\ln\left(\frac{V_{\mathrm{end}}^{1/4}}{\mpl}\right)\,.\label{eq:efolds_final_w1}
\end{equation}

\section{Results}

To find the energy scale at the end of inflation we can first substitute
$n_{s}$ from Eq.~\eqref{aN} into Eq.~\eqref{V0As} to obtain
\begin{equation}
\frac{V_{0}}{\mpl^{4}}=\frac{12\pi^{2}\alpha A_{\mathrm{s}}}{N_{*}^{2}}\,e^{\alpha/N_{*}}\,,\label{V0fin}
\end{equation}
which is independent of $\kappa$. Using that %
\mbox{%
$\alpha<N_{*}$%
} and 
\mbox{%
$A_{\mathrm{s}}=(2.208\pm0.075)\times10^{-9}$%
} (cf.~Eq.~\eqref{As-val}), this always gives 
\begin{equation}
V_{0}^{1/4}\simeq2\times10^{-3}\alpha^{1/4}\mpl\,.\label{V0}
\end{equation}
Using this value to find %
\mbox{%
$V_{{\rm end}}=V_{0}e^{-\sqrt{2\alpha}}$%
} we see 
\begin{equation}
V_{\mathrm{end}}^{1/4}=
2\times10^{-3}\alpha^{1/4}e^{-\sqrt{\alpha/8}}\mpl\sim10^{-3}\mpl\,,\label{eq:Vend}
\end{equation}
where %
\mbox{%
$\alpha\in[1.5,10]$%
} (see also Fig.~\eqref{fig:VN}). This is close to the scale of
a grand unified theory (GUT) as expected.

Rearranging Eq.~\eqref{Treh} we readily obtain 
\begin{equation}
\frac{V_{{\rm end}}^{1/4}}{T_{{\rm reh}}}=\left[\frac{\pi^{2}g_{*}}{10}\left(\frac{\kappa}{\sqrt{2\alpha}}\frac{\phi_{{\rm ESP}}}{\mpl}\right)^{\sqrt{6\alpha}}\right]^{1/4}\,.\label{VendTreh}
\end{equation}
Also, using %
\mbox{%
$V_{{\rm end}}=V_{0}e^{-\sqrt{2\alpha}}$%
} and Eq.~\eqref{V0fin} we find 
\begin{equation}
\frac{V_{{\rm end}}^{1/4}}{\mpl}=\frac{\left(12\pi^{2}\alpha A_{\mathrm{s}}\right)^{1/4}}{\sqrt{N}_{*}}\exp\left(\frac{\alpha}{4N_{*}}-\sqrt{\frac{\alpha}{8}}\right)\,.\label{Vendmpl}
\end{equation}
Combining eqs.~\eqref{VendTreh} and \eqref{Vendmpl} with Eq.~\eqref{eq:efolds_final_w1}
and after some algebra we end up with 
\begin{equation}
N_{*}\simeq56.3+\sqrt{\frac{\alpha}{24}}\,\ln\left(\frac{\kappa}{\sqrt{2\alpha}}\frac{\phi_{{\rm ESP}}}{\mpl}\right)\,,\label{N*}
\end{equation}
where we took %
\mbox{%
$\exp\left(\frac{\alpha}{4N_{*}}-\sqrt{\frac{\alpha}{8}}\right)\sim1$%
}.

The value of $\kappa$ is determined by the necessity for the residual
potential energy of $\varphi$ to act as dark energy at late times
(see Eq.~\eqref{k-Vs}). As such 
\begin{equation}
V(\varphi_{\mathrm{ESP}})=V_{0}\exp\left(-\kappa e^{\frac{\varphi_{\mathrm{ESP}}}{\sqrt{\alpha}\mpl}}\right)\simeq10^{-120}\mpl^{4}\,,
\end{equation}
which rearranges to 
\begin{equation}
\kappa\approx244\,e^{\frac{\varphi_{\mathrm{ESP}}}{\sqrt{\alpha}\mpl}}\label{kappaexp}
\end{equation}
(cf. Eq.~\eqref{kappa-b}) Using this, Eq.~\eqref{N*} is written
as 
\begin{equation}
N_{*}\simeq56.3+\sqrt{\frac{\alpha}{24}}\left[\ln\left(\frac{244}{\sqrt{2\alpha}}
\right)+\frac{2\varphi_{{\rm ESP}}}{\sqrt{\alpha}\mpl}\right]\,,\label{N*fin}
\end{equation}
where we used %
\mbox{%
$\ln\left(\frac{\phi_{{\rm ESP}}}{\mpl}\right)=\frac{\varphi_{{\rm ESP}}}{\sqrt{\alpha}\mpl}$%
}, according to Eq.~\eqref{phivarphi}.

We know that the pole in the non-canonical field potential, $\phi=0$
is transposed to $\varphi=-\infty$ with our field redefinition, generating
a plateau in the model which provides the slow-roll regime for inflation.
The value of $\kappa$ effectively shifts the position of the edge
of the plateau, which explains why the value of $\varphi_{\mathrm{ESP}}$
differs for each $\kappa$ value in this equation.

\begin{figure}
\begin{centering}
\includegraphics[scale=0.5]{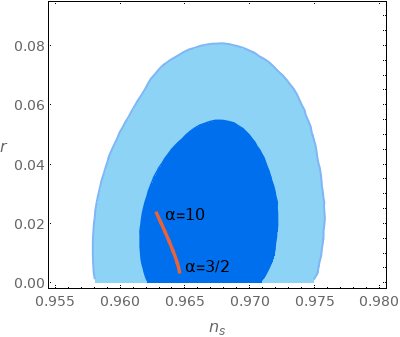}
\par\end{centering}
\caption{\label{fig:Planck}The $1\sigma$ and $2\sigma$ contours of the Planck
constraints from Ref.~\citep{Akrami:2018odb}. The red curve is the
prediction of our model assuming Eq.~\eqref{dphi-Treh}.}
\end{figure}

It is straightforward to obtain the inflationary observables $n_{s}(\alpha)$
and $r(\alpha)$ using eqs.~\eqref{spectral} and \eqref{ratio}
with Eq.~\eqref{N*fin}. As an indicative choice we consider %
\mbox{%
$\varphi_{{\rm ESP}}=0$%
}. In this case, the results are shown in Table~\ref{table} and depicted
in Fig.~\ref{fig:Planck}. As evident, there is excellent agreement
with the Planck results \citep{Akrami:2018odb}.

\begin{table}
\begin{tabular}{|c||c|c|c|}
\hline 
$\alpha$  & $N_{*}$  & $n_{s}$  & $r$ \tabularnewline
\hline 
\hline 
3/2  & 57.5  & 0.965  & 0.0024 \tabularnewline
\hline 
10  & 58.9  & 0.963  & 0.023\tabularnewline
\hline 
\end{tabular}\caption{Inflationary observables taking %
\mbox{%
$\varphi_{{\rm ESP}}=0$%
}.}
\label{table} 
\end{table}

Remarkably, the values of the inflationary observables do not change
much when varying $\varphi_{{\rm ESP}}$. For example, taking %
\mbox{%
$\varphi_{{\rm ESP}}=\mpl$%
} adds %
\mbox{%
$2/\sqrt{24}\simeq0.4$%
} to the value of $N_{*}$, so that %
\mbox{%
$N_{*}(3/2)=57.9$%
} (%
\mbox{%
$N_{*}(10)=59.3$%
}). Yet, the resulting values of $n_{s}$ and $r$ remain virtually
unchanged, given by the same values shown in Table~\ref{table}.
Thus, our results are robust and only very weakly dependent on the
location of the ESP (value of $\varphi_{{\rm ESP}}$), which means
that no tuning is required to match the observations.

This is less so with the value of $\kappa$. Indeed, from Eq.~\eqref{kappaexp}
we see that setting $\varphi_{\mathrm{ESP}}=0$ requires 
\begin{equation}
\kappa\approx244\,,\label{k1}
\end{equation}
to obtain the correct energy density for dark energy. If instead we
have $\varphi_{\mathrm{ESP}}=\mpl$, with $\alpha=10$, we have 
\begin{equation}
\kappa\approx335\,.\label{k2}
\end{equation}
In all cases however, we see that %
\mbox{%
$\kappa={\cal O}(100)$%
}, which means that the inflaton field in the exponent of the potential
in Eq.~\eqref{L-org} is suppressed by the GUT scale %
\mbox{%
$\mpl/\kappa\sim10^{16}\,$GeV%
}.

\section{Conclusions}

We have analysed a new model of quintessential inflation, inspired
by supergravity and superstrings. The inflaton field features a runaway
potential with a kinetic pole at the origin that generates the inflationary
plateau. After the field rolls over the edge of this plateau, it becomes
kinetically dominated, driving a period of kination. The rapid roll
of the inflaton is halted, when it crosses an enhanced symmetry point
(ESP), where its kinetic density is transferred to the generation
of the thermal bath of the hot big bang, through interaction with
the Peccei-Quinn (PQ) field. Thereby, the roll of the inflaton is
stopped before it travels over super-Planckian distances in field
space, which would otherwise undermine the validity of the scalar
potential. Trapping the field at the ESP not only reheats the Universe
but also ensures that the field becomes heavy and does not give rise
to the 5th force problem, which typically plagues quintessence models.
The residual potential density of the field can explain dark energy
without resorting to a non-zero value of the cosmological constant.
Another aspect of our model which significantly differs from other
quintessential inflation models in the literature is that radiation
production occurs at reheating and not before, meaning that there
is no subdominant thermal bath during kination.\footnote{except due to
  gravitational particle production \cite{Ford:1986sy}\cite{Chun:2009yu}.}

We have studied in detail how the kinetic density of the inflaton
is transferred to radiation through the tachyonic and parametric resonant
production of PQ particles. Coupling the inflaton to the PQ field
is aligned with the economy philosophy of quintessential inflation,
in that no arbitrary new field is introduced by hand to interact with
the inflaton field responsible for both inflation and dark energy,
but the field considered (the PQ field) is already envisaged by beyond
the standard model physics to account for the strong CP problem of
QCD and for the dark matter in the Universe. Moreover, in our model,
the interaction between the inflaton and the PQ field ensures that
the PQ symmetry is restored during inflation. As a result, the axion
field does not exist during inflation, so it does not obtain a superhorizon
spectrum of perturbations of its expectation value. This means that
there is no issue of axion isocurvature perturbations, which can otherwise
be a concern when considering axionic dark matter.\footnote{At the breaking of the PQ symmetry the Kibble mechanism may give rise
to axionic cosmic strings, which however are harmless since their
tension is at the PQ scale %
\mbox{%
$G\mu\sim(f/\mpl)^{2}\sim10^{-12}$%
}, so they do not introduce any dangerous signals in the CMB.}

Our model manages to account for observations {with natural values of the
  model parameters.}
The inflationary observables obtained (see Table~\ref{table}) are
in excellent agreement with the Planck satellite findings and are
rather robust, in that they do not significantly depend on the location
of the ESP down the runaway inflaton direction. This is not surprising,
given that the inflationary plateau is generated by the presence of
a kinetic pole. For dark energy, the strength of the exponential characterising
the potential slope only implies that the inflaton is suppressed by
the GUT scale.

All in all, we have presented a new quintessential inflation model
which successfully accounts for inflation and dark energy and may well have a
basis in fundamental physics.


\section*{Acknowledgements}

Consortium for Fundamental Physics under the STFC Grant No. ST/L000520/1.
M.K. is supported by the Communidad de Madrid "Atracci\'{o}n
de Talento investigador" grant 2017-T1/TIC-5305 and
MINECO (Spain) project \{FIS2016-78859-P(AEI/FEDER, UE)\}. C.O. is supported
by the Faculty of Science and Technology of Lancaster University.


\bibliographystyle{aipnum4-1}
\bibliography{QIwithTrap}

\end{document}